\documentclass[prd,aps,nofootinbib,floats,letterpaper,floatfix,superscriptaddress,groupedaddress,eqsecnum,showpacs,twocolumn]{revtex4}

\usepackage{graphicx,epsf, epsfig, amssymb}
\usepackage{amsfonts,amsmath,amssymb,amsthm}
\usepackage{bm}
\usepackage{dcolumn}
\usepackage{latexsym}
\usepackage{rotating}
\usepackage[dvips]{color}
\usepackage{longtable}

\def\ii{{\rm i}}

\begin{document}

\title{Gravitational signature of Schwarzschild black holes\\ in dynamical Chern-Simons gravity}

\author{C. Molina} \email{cmolina@usp.br} \affiliation{Escola de Artes, Ci\^{e}ncias e Humanidades, Universidade de
  S\~{a}o Paulo,\\
  Avenida Arlindo Bettio 1000, 03828-000 S\~{a}o Paulo, SP, Brazil}

\author{Paolo Pani} \email{paolo.pani@ca.infn.it} \affiliation{Dipartimento di
  Fisica, Universit\`a di Cagliari, and INFN sezione di Cagliari, Cittadella
  Universitaria 09042 Monserrato, Italy}

\author{Vitor Cardoso} \email{vitor.cardoso@ist.utl.pt}
\affiliation{Centro
  Multidisciplinar de Astrof\'{\i}sica - CENTRA, Dept. de F\'{\i}sica,
  Instituto Superior T\'ecnico, Av. Rovisco Pais 1, 1049-001 Lisboa, Portugal} 
 \affiliation{Department of Physics and Astronomy, The University of
  Mississippi, University, MS 38677-1848, USA}

\author{Leonardo Gualtieri} \email{leonardo.gualtieri@roma1.infn.it} \affiliation{Dipartimento di Fisica, Universit\`a di Roma
``Sapienza'' and Sezione INFN Roma1, P.A. Moro 5, 00185, Roma, Italy}


\begin{abstract}
  Dynamical Chern-Simons gravity is an extension of General Relativity in which the gravitational field is coupled to a
  scalar field through a parity-violating Chern-Simons term. In this framework, we study perturbations of spherically
  symmetric black hole spacetimes, assuming that the background scalar field vanishes. Our results suggest that these
  spacetimes are stable, and small perturbations die away as a ringdown. However, in contrast to standard General
  Relativity, the gravitational waveforms are also driven by the scalar field. Thus, the gravitational oscillation modes
  of black holes carry imprints of the coupling to the scalar field. This is a smoking gun for Chern-Simons theory and
  could be tested with gravitational-wave detectors, such as LIGO or LISA. For negative values of the coupling constant,
  ghosts are known to arise, and we explicitly verify their appearance numerically. Our results are validated using both
  time evolution and frequency domain methods. 
\end{abstract}


\maketitle

\section{Introduction}
In Chern-Simons gravity \cite{Deser:1982vy,Lue:1998mq,Jackiw:2003pm} the Einstein-Hilbert action is modified by adding a
parity-violating Chern-Simons term, which couples to gravity via a scalar field. This correction could explain several
problems of cosmology \cite{Weinberg:2008mc,GarciaBellido:2003wd,Alexander:2004xd,Alexander:2004us,Konno:2008np}. Furthermore, a
Chern-Simons term arises in many versions of string theory \cite{Polchinski:1998rr} and of loop quantum gravity
\cite{Ashtekar:1988sw,Taveras:2008yf,Mercuri:2009zt}, and Chern-Simons gravity can be recovered by truncation of low
energy effective string models \cite{Smith:2007jm,AD08}.

When Chern-Simons gravity was first formulated, the scalar field was considered as a prescribed function. Later on, it
was understood that this {\it a priori} choice is not really motivated (see the discussion in
Ref.~\cite{Yunes:2009hc}). Then, dynamical Chern-Simons (DCS) gravity has been formulated \cite{Smith:2007jm}, in which
the scalar field is treated as a dynamical field.

Since DCS gravity has a characteristic signature (the Chern-Simons term violates parity), there is the exciting prospect
of testing its predictions against astrophysical observations. This has motivated a large body of work on the subject
(for a review on DCS gravity and its astrophysical consequences see Ref.~\cite{Alexander:2009tp}).  In this context, the
study of black hole (BH) perturbations is very promising, since astrophysical black holes are probably the most
appropriate objects to probe the strong field regime of General Relativity \cite{Sopuerta:2009iy}.

The first study of BH perturbations in the context of DCS gravity has been carried out in Ref.~\cite{Yunes:2007ss},
where it was found that, if the background solution contains a (spherically symmetric) scalar field, polar and axial
perturbations of DCS BHs are coupled, and the equations describing them are extremely involved. Recently, in
Ref.~\cite{Cardoso:2009pk} (hereafter, Paper I), some of us found that, when the background scalar field vanishes, polar
and axial gravitational perturbations of a Schwarzschild BH decouple, and only axial parity perturbations are affected
by the Chern-Simons scalar field. We also found that under this assumption the gravitational and scalar perturbations
are described by a coupled system of two second order ordinary differential equations (ODEs). The numerical integration
of this system to find the quasi-normal modes (QNMs) of Schwarzschild DCS BHs is challenging, due to the same asymptotic
divergence which prevented for many years the numerical computation of QNMs for Schwarzschild BHs
\cite{Chandrasekhar:1975zz,Nollert:1999ji,Berti:2009kk,Ferrari:2007dd}. Therefore, in Paper I the QNMs of Schwarzschild
DCS BHs were not investigated thoroughly. It is remarkable that there are very few studies of this kind of system, i.e.,
QNMs described by coupled ODEs (one interesting work is presented in Ref. \cite{Seahra:2005}). In Paper I we also
claimed that Schwarzschild DCS BHs are unstable for a specific range of the parameters of the theory. This result was
the consequence of a sign error in the derivation of the perturbation equations; on the contrary, as we discuss in this
paper, there is strong evidence that these spacetimes are stable.

In this paper we complete the study of Schwarzschild DCS perturbations, performing a thorough numerical analysis of the
perturbation equations. We employ two different -- and completely independent -- numerical approaches: time evolution
and a formulation of the frequency domain approach \cite{Watanabe:1980} which has never been applied before to the study
of instability in black hole spacetimes. The results of the two independent methods agree very well, typically within an
accuracy of $0.1\%$, validating each other.

The main result we find is that any perturbation decays at late-time as a damped sinusoid. This is known as the ringdown
phase, where the black hole radiates all excess hairs in its lowest QNMs \cite{Berti:2009kk,Ferrari:2007dd}. What is new
here, and with important implications for tests of DCS gravity, is that the gravitational sector has two distinct sets
of QNMs. For large values of the constant $\beta$ (associated to the dynamical coupling of the scalar field), these two
sets coincide with the usual gravitational QNMs and scalar field QNMs of General Relativity. This result enables simple,
yet fundamental tests on DCS gravity. By measuring (or not) these two different modes, one could effectively constrain
DCS gravity through gravitational-wave observations. For instance, detection of ringdown modes with a signal-to-noise
ratio $\gtrsim 6$ (feasible with both the Earth-based LIGO and the space-based detector LISA), could allow one to test
DCS gravity if the mass of the BH is known, for instance through observations of the inspiral phase of black hole
binaries. For signal-to-noise ratios $\gtrsim 150$ one could be able to discriminate between DCS gravity and standard
General Relativity without any further knowledge of the BH parameters.
\subsection*{A summary of our results}

For the reader wishing to skip the technical details of the rest of the paper, the following is a brief summary of our
results.

\begin{itemize}

\item[(i)] Two complementary numerical methods were developed and employed. They are completely independent and their
  concordance is very good.

\item[(ii)] For small values of the coupling constant ($M^4 \beta\lesssim0.5$), the perturbative dynamics is
  characterized by a stable exponentially decaying phase. The intermediate late time evolution is dominated by
\begin{equation}
\Phi(t,r_{\rm fixed}) = e^{\omega_{\rm no} \, t}
\left(\begin{array}{c}
  a\\
  b
\end{array} \right)
\end{equation}
with $\textrm{Re}[\omega_{\rm no}] = 0$ and $\textrm{Im}[\omega_{\rm no}] < 0$ (with our sign conventions, a QNM is stable if
$\textrm{Im}[\omega]<0$). Our results for the non-oscillatory frequency values are compatible with the expression:
\begin{equation}
\omega_{\rm no} = - 0.04024 (M^4 \beta)^{0.44} \ell 
\left(1 + \frac{2.0953}{\ell} - \frac{3.4460}{\ell^2} \right)\,.
\end{equation}

\item[(iii)] For intermediate values of $M^4 \beta$, field evolution is dominated by a stable oscillatory phase. We have
  detected two oscillatory modes, named here ``gravitational'' and ``scalar'' modes. Although the time profiles of the
  gravitational perturbation $\Psi$ and of the scalar field $\Theta$ are usually different, they consist on different
  superpositions of the {\it same} modes.

\item[(iv)] In the $\beta\to\infty$ limit, these ``gravitational'' and ``scalar'' branches coincide with actual
  gravitational and scalar modes of Schwarzschild BHs in General Relativity. In this regime, we report that for
  $\ell=2$, we find that the gravitational perturbation oscillates with a combination of the two modes
\begin{eqnarray}
M \omega_{\rm grav} (\Psi)&=&0.3736 - {\rm i} \,\, 0.08899\,,\\
M \omega_{\rm sc} (\Psi)&=&0.4837 - {\rm i} \,\, 0.09671\,.
\end{eqnarray}
These numbers correspond to the lowest mode of pure gravitational and scalar quasi-normal frequencies in Einstein's
theory \cite{Berti:2009kk}. The scalar field perturbation, instead, oscillates with the mode $\omega_{\rm sc}$ only. This
behavior can be easily understood by looking at the form of the equations in this limit.

\item[(v)] At late times, the field decays with a power-law tail, as $t^{- (2\ell + 3)}$. The tails do not depend on
  $\beta$ or $M$. Note that the same behavior characterizes Schwarzschild BHs \cite{CPM78}, implying that a
  gravitational-wave observation of the tail would not be able to discriminate DCS gravity from General Relativity.
 
\item[(vi)] An extensive investigation of BH oscillations, performed using two different numerical approaches, only
  yields stable modes, either oscillating or non-oscillating. This gives strong indications that Schwarzschild BHs in
  DCS modified gravity are stable against axial and polar perturbations.
 
\item[(vii)] We also discuss how the inclusion of a non-vanishing scalar potential in the Lagrangian affects the QNM
  spectrum.  We focus on potentials of the form
\begin{equation}
V(\vartheta)=m^2\vartheta^2+{\cal O}(\vartheta^3) \label{potential}
\end{equation}
and find that in the $\beta\to\infty$ limit this inclusion only affects the scalar branch of QNMs, while the
gravitational branch is unaltered. When $M^4\beta\lesssim100$, also the gravitational sector is affected by the
scalar potential.
\end{itemize}

The paper is organized as follows. In Section \ref{eqns} we briefly review the derivation of the perturbation equations
in DCS gravity. In Section \ref{num1} we describe the time
domain and frequency domain numerical approaches that we have employed to solve the perturbation equations. In Section
\ref{results} we present our results in the time and frequency domains.  In Section \ref{tests}, a possible
observational signature of DCS gravity is discussed. Implications and final remarks are presented in Section
\ref{concl}.

In Appendix \ref{sec:negative_beta} we discuss ghost-like instabilities arising when the wrong sign of the kinetic term
in the action is chosen, i.e. when $\beta<0$ in Eq.~\ref{action} below.

\section{Perturbation equations and dynamical stability}\label{eqns}
The action of DCS gravity is \cite{Yunes:2009hc}
\begin{eqnarray}
&&S=\kappa\int d^4x\sqrt{-g}R+\frac{\alpha}{4}\int d^4x\sqrt{-g}
\vartheta\,^*RR\nonumber\\
&&-\frac{\beta}{2}\int d^4x\sqrt{-g}\left[
g^{ab}\nabla_a\vartheta\nabla_b\vartheta+V(\vartheta)\right]+S_{\rm mat}\,.\label{action}
\end{eqnarray}
where $\vartheta$ is the scalar field and
\begin{equation}
^*RR=\frac{1}{2}R_{abcd}\epsilon^{baef}R^{cd}_{~~ef}\,.
\end{equation}
We use geometrical units $c=G=1$ so that $\kappa=1/16\pi$. Furthermore, we neglect $V(\vartheta)$ (this assumption will
be relaxed in Section \ref{massterm}), and consider the vacuum solutions ($S_{\rm mat}=0$). The equations of motion are
\begin{eqnarray}
R_{ab}&=&-16\pi \alpha C_{ab}+8\pi\beta\vartheta_{,a}\vartheta_{,b}\label{eqE1}\\
\Box\vartheta&=&-\frac{\alpha}{4\beta}\,^*RR \label{eqE2}
\end{eqnarray}
where
\begin{equation} 
C^{ab}=\vartheta_{;c}\epsilon^{cde(a}\nabla_eR^{b)}_{~~d}
+\vartheta_{;dc}\,^*R^{d(ab)c}\,.\label{Ctensor}
\end{equation} 
In a spherically symmetric background, $^*RR=0=C^{ab}$ and Eqs. (\ref{eqE1}), (\ref{eqE2}) reduce to usual Einstein
gravity minimally coupled to a scalar field
\begin{eqnarray}
R_{ab}=8\pi \vartheta_{,a}\vartheta_{,b}\,,\quad \Box\vartheta=0\,.
\end{eqnarray}
No-hair theorems \cite{Bekenstein:1972ny} state that the Schwarzschild solution, with vanishing scalar field, is the
only static spherically symmetric solution of the equations above. 
We then consider perturbations of a Schwarzschild BH with a vanishing background scalar field. We expand the gravitational
perturbations in tensor spherical harmonics, building the Zerilli and Regge-Wheeler functions. The scalar field is
expanded in scalar spherical harmonics as
\begin{equation}
\vartheta=\frac{\Theta^{\ell m}}{r}Y^{\ell m}e^{-{\rm i}\omega t}\,.\label{expscalar}
\end{equation}
Eq.~(\ref{eqE1}) implies (see Paper I) that polar parity gravitational perturbations (described by the Zerilli function)
are not affected by the Chern-Simons scalar, and then the corresponding QNMs are the well-known modes of Schwarzschild
BHs. Axial parity gravitational perturbations $\Psi^{\ell m}={\rm i}Q^{\ell m}/\omega$ (where $Q^{\ell m}$ is the
Regge-Wheeler function, defined as in Paper I) are instead coupled with the scalar field. From here onwards, we will
drop the $^{\ell m}$ superscripts.

Eqs.~(\ref{eqE1}),
(\ref{eqE2}) reduce to the following set of coupled ordinary differential equations for the perturbations
$\Theta(r)$ and $\Psi(r)$, in terms of which one can completely characterize the axial parity metric
perturbations and the scalar field respectively:
\begin{widetext}
\begin{eqnarray}
\frac{d^2}{dr_\star^2}\Psi+\left\{ \omega^2-f\left[ \frac{\ell(\ell+1)}{r^2}-\frac{6M}{r^3}\right]\right\}\Psi&=&
\frac{96\pi\,Mf}{r^5}\alpha \Theta\,,\label{eqq2}\\
\frac{d^2}{dr_\star^2}\Theta+\left\{\omega^2-f\left[\frac{\ell(\ell+1)}{r^2}
\left(1+\frac{576\pi M^2\alpha^2}{r^6\beta}\right)
+\frac{2M}{r^3}\right]\right\}\Theta&=&f\frac{(\ell+2)!}{(\ell-2)!}\frac{6 M\alpha}{r^5\beta}
\Psi\label{eqpsi2}
\end{eqnarray}
\end{widetext}
with $f(r)=1-2M/r$ and $r_{\star}\equiv r+2M\ln\left(r/2M-1\right)$. Note that third time-derivatives (i.e. terms
proportional to $\omega^3$) do not arise in the perturbation equations above (although they are generally expected from
Eqs.~(\ref{eqE1})-(\ref{eqE2})) because of the vanishing of the background Ricci tensor in Eq.~(\ref{Ctensor}).
Therefore the Schwarzschild background does not suffer from problems related to ill-posedness of the theory, the
so-called Ostrogradski instability (see Refs.~\cite{Alexander:2009tp,Yunes:2009ch}). We also remark that the instability
found in Paper I for $\beta M^4\lesssim2\pi$ was an artifact of a wrong sign in the definition of $^*RR$, that has yield
a change in the sign of $\beta$ in the perturbation equations. This is equivalent to consider the equations of the DCS
theory with $\beta<0$, which is indeed expected to be unstable, as discussed in Appendix \ref{sec:negative_beta}.
\subsection*{Re-scaling and the General Relativity limit}
Under the replacement $\beta\to \alpha^2 \beta$ and $\Theta\to \Theta/\alpha$, one can set $\alpha=1$ in the
perturbation equations (\ref{eqq2}) and (\ref{eqpsi2}), which we will hereafter assume. Indeed, as discussed in
\cite{Alexander:2009tp}, the parameters of the theory are redundant, and it is always possible to fix one of them.

We remark, however, that there is a subtle formal difference between the theory with $\alpha,\beta$ and the theory with
$\alpha=1$. Indeed, the General Relativity limit of the former is obtained by taking $\beta\to \infty$ and $\alpha \to
0$; the General Relativity limit of the latter is obtained by taking $\beta\to\infty$ and by considering the solutions
with $\Theta\equiv0$. In other words, once we fix $\alpha=1$, General Relativity is not simply a limit of the DCS
theory: it is a particular subset of the solution space of the $\beta\to\infty$  limit of the theory.

\section{Numerical approach}\label{num1}

\subsection{Time domain evolution}

The system (\ref{eqq2}), (\ref{eqpsi2}) can be written as
\begin{equation}
\left(-\frac{\partial^{2}}{\partial t^{2}}+\frac{\partial^{2}}{\partial r_{\star}^{2}}\right)\Phi=V\,\Phi
\end{equation}
where we have defined
\begin{equation}
\Phi=\left(\begin{array}{c}
 \Psi\\
 \Theta
\end{array} \right)\,,\quad V=\left(\begin{array}{cc}
 V_{11} & V_{12}\\
 V_{21} & V_{22}
\end{array} \right)\,,\label{matrix_eq}
\end{equation}
and the elements of the matrix potential $V$ are given by
\begin{eqnarray}
V_{11}&=&f(r)\left[\frac{\ell(\ell+1)}{r^{2}}-\frac{6M}{r^{3}}\right]\,,\\
V_{12}&=&f(r)\frac{96\pi M}{r^{5}}\,,\\
V_{21}&=&f(r)\frac{6M(\ell+2)!}{\beta(\ell-2)!}\frac{1}{r^{5}}\,,\label{V21}\\
V_{22}&=&f(r)\!\left[\frac{\ell(\ell\!+\!1)}{r^{2}}\!\left(\!1\!+\!\frac{576\pi M^{2}}{\beta\,r^6}\!\right)
\! +\! \frac{2M}{r^{3}} \right].\label{V22}
\end{eqnarray}
Using the light-cone variables $u=r_{\star}-t$ and $v=r_{\star}+t$ one can write 
\begin{equation}
\label{eqm_uv}
4\frac{\partial^{2}}{\partial u\partial v}\Phi=-V\,\Phi\,.
\end{equation}

A discretized version of Eq. (\ref{eqm_uv}) is
\begin{eqnarray}
&&\Phi(N) - \Phi(E) - \Phi(W) + \Phi(S)\nonumber\\
&& =\frac{\Delta u\Delta v}{8}  \, V(S) \,\left[\Phi(E)+\Phi(W)\right]\,,
\label{disc_1}
\end{eqnarray}
where the points $N, E, W, S$ are defined as follows: $N = (u + \Delta, v + \Delta)$, $W = (u + \Delta, v)$, $E =
(u, v + \Delta)$ and $S = (u,v)$. With the expression (\ref{disc_1}), the region of interest in the $u-v$ plane is
covered, using the value of the field at three points in order to calculate the fourth one. As the integration proceeds,
the values of $\Psi (t, r_{\rm fixed})$ are extracted \cite{Wang:2000dt,Wang:2004bv}.

The initial data consist of the expressions on the sub-manifolds $(u>0,v=0)$ and $(u=0,v>0)$ for the vector
\begin{equation}
\Phi(u,v)= \left(\begin{array}{c}
   \Psi(u,v)\\
   \Theta(u,v)
\end{array} \right)\,.
\end{equation}
For most of the numerical evolutions presented here the initial data have the form
\begin{eqnarray}
\Phi(u,0)&=& \left(\begin{array}{c}
   0\\
   0
\end{array} \right)\,,\\
\Phi(0,v)&=&e^{-(v-v_{c})^{2}/2\sigma}
\left(\begin{array}{c}
  1\\
  1
\end{array} \right)\,,\label{gauss}
\end{eqnarray}
with $v_{c} = 10.0$ and $\sigma = 1.0$. 

From results on BH oscillations in General Relativity \cite{Nollert:1999ji} we expect that the main
characteristics of the time-evolution profiles (after a transient initial regime) are insensitive to the choice of the
initial data, provided that they are localized. To check if this actually occurs in the present case, and rule out any
eventual influence of initial data on late time results, we have considered different choices for the initial data:
\begin{itemize}
\item Gaussian initial data
\begin{equation}
\Phi(0,v) =
\left(\begin{array}{c}
 A_{1}\, e^{-(v-v_{c1})^{2}/2\sigma_1}\\
 A_{2}\, e^{-(v-v_{c2})^{2}/2\sigma_2}
\end{array} \right)\,,\quad
\Phi(u,0)= \left(\begin{array}{c}
   0\\
   0
\end{array} \right)\,.\label{initdata}
\end{equation}
The initial $v$-functions are localized, with different peaks for the $\Psi$ and $\Theta$ components. Although
strictly speaking they do not have compact supports, they are (numerically) zero far away from the peaks.
\item Compact support pulses
\begin{equation}
\Phi(0,v) =
f(v)
\left(\begin{array}{c}
   1\\
   1
\end{array} \right)\,,\quad
\Phi(u,0)= \left(\begin{array}{c}
   0\\
   0
\end{array} \right)\,.
\end{equation}
We have chosen two different functions $f=f_1(v),f_2(v)$. The first choice corresponds to 
\begin{equation}
f_1(v) = \left[ 4 \frac{\left(v - v_{2}\right)\left(v - v_{1}\right)}
   {\left(v_{2} - v_{1}\right)^2} \right]^8\,,\, v_{1}< v < v_{2}
\end{equation}
and zero elsewhere. This is a localized and smooth pulse with a compact support. Our second choice corresponds to
\begin{equation}
f_2(v) = 1\,,\qquad v_{1}< v < v_{2}\,,
\end{equation}
and zero elsewhere. It is a localized but not continuous pulse with a compact support.
\end{itemize}
We have verified that the numerical results (after a transient regime) do not depend on the initial data.

\subsection{Iteration scheme in the frequency domain}\label{sec:method}
We now present an alternative, and complementary, numerical method, which is an application of Newton's iteration scheme
to the shooting method \cite{Watanabe:1980}.

Let us define $\omega_0$ as the trial eigenfrequency of the eigenvalue problem defined by Eqs.~(\ref{eqq2}),
(\ref{eqpsi2}). The corresponding solutions $\Psi_0$ and $\Theta_0$ satisfy the following set of equations
\begin{eqnarray}
\Psi_0''(r_\star)+(\omega^2-V_{11})\Psi_0(r_\star)&=&V_{12}\Theta_0(r_\star)\,,\label{eqm1}\\
\Theta_0''(r_\star)+(\omega^2-V_{22})\Theta_0(r_\star)&=&V_{21}\Phi_0(r_\star)\label{eqm2}\,,
\end{eqnarray}
and it is hereafter understood that all these quantities are evaluated at the trial frequency
$\omega_0$. In order to compute QNMs we require the following boundary conditions
\begin{equation}
\Phi_0(\pm\infty)=\left(\begin{array}{c}
  \Psi_0(\pm\infty)\\
  \Theta_0(\pm\infty)
\end{array} \right) \sim \left(\begin{array}{c}
  A_\pm\\
  B_\pm
\end{array} \right)e^{\pm i\omega r_\star}\label{BC}\,.
\end{equation}
When $\textrm{Im}[\omega]<0$ Eq.~(\ref{BC}) defines (stable) QNMs, while when $\textrm{Im}[\omega]>0$ we have
``bound-state-like'' boundary conditions, i.e. $\Phi_0 \to 0$ at $r_{\star}\to\pm \infty$ (see Paper I) and the
corresponding modes are unstable. The numerical method described in the rest of this section is capable to find both
stable and unstable modes.

The idea is to ``shoot'' from each of the boundaries to a matching point where the wave functions and their derivatives
are required to be continuous. In general, $\omega_0$ is not the true eigenfrequency, and one of the continuity
equations for $\Psi_0$ and $\Theta_0$ is not satisfied. Without loss of generality, we can choose either $\Theta_0$ or
$\Theta_0'$ to be the function which does not satisfy the continuity condition. Moreover we consider the matching point
to be at $r_\star=0$. Namely we assume that
\begin{equation}
\left[\left[\Psi_0\right]\right]=\left[\left[\Psi_0\right]\right]=\left[\left[\Theta_0\right]\right]=0\,,
\qquad \left[\left[\Theta_0'\right]\right]\neq0\,,\label{eqcont}
\end{equation}
where we define $\left[\left[\dots\right]\right]$ as the difference between the limits of the corresponding quantity as
$r_\star\to 0_\pm$. We checked that our numerical results do not depend on the choice of the matching point within a
wide range around $r_\star=0$.  We perform two integrations: one starting at $+\infty$ (numerically, at
$r_\star=r_\star^{(1)}\gg M$) inward to $r_\star=0$, and the other one starting at $-\infty$ (numerically, at
$r_\star=r_\star^{(2)}\ll-M$) outward to $r_\star=0$. At both infinities, we expand solution in series as follows
%
\begin{equation}
 \left(\begin{array}{c}
  \Psi\\
  \Theta
\end{array} \right) 
\sim \left\{
\begin{array}{ll}
\displaystyle 
\left(\begin{array}{c}
  A_H\\
  B_H
\end{array} \right)e^{-i\omega r_\star}\left[1+\sum_{n=1}^N\left(\begin{array}{c}
  a_H^{(n)}\\
  b_H^{(n)}
\end{array}\right)(r-2M)^n\right], \\
\\
\displaystyle
\left(\begin{array}{c}
  A_\infty\\
  B_\infty
\end{array} \right)e^{i\omega r_\star}\left[1+\sum_{n=1}^N\left(\begin{array}{c}
  a_\infty^{(n)}\\
  b_\infty^{(n)}
\end{array}\right)r^{-n}\right],  
\end{array}
\right.\label{series}
\end{equation}
%
at $r_\star\ll -M$ and at $r_\star\gg M$, respectively. In computing QNMs this way it is important to choose appropriate
values of numerical infinities, because numerical instabilities may arise by considering too large values for
$r_\star^{(1)}$ and $r_\star^{(2)}$ \cite{Chandrasekhar:1975zz}. In fact at both infinities the general solution will be
a mixture of exponentially growing and exponentially suppressed modes and (in order to compute QNMs) we must select pure
exponentially growing modes. Problems arise when too large values for $r_\star^{(1)}$ and $r_\star^{(2)}$ are chosen,
because in that case contributions from unwanted exponentially suppressed modes can be significant after the
integration, due to numerical errors. This problem can be circumvented by choosing small enough values of numerical
infinities, say $|r_\star^{(i)}|\sim10M$, and by considering large enough order of series expansion $N$, say
$N\gtrsim10$. In this way, though $\sim 10M$ is not very large (typically, for the modes we find, $|10M\omega|\sim 3-5$) the series well
approximates the correct solution.
This problem does not arise in the computation of unstable modes (see Appendix \ref{sec:negative_beta}), since in that
case we simply impose Dirichlet conditions at both infinities.

In order to obtain solutions satisfying continuity conditions~(\ref{eqcont}) at the matching point, we compute two
\emph{linear independent} solutions and we construct an appropriate linear combination of them, which satisfies the
required conditions. The first solution is obtained by choosing $A=1$ and a generic value $B=B_0$ in the series
expansion, whereas the second solution is obtained by choosing $B=1$ and a generic value of $A=A_0$. We shall denote the
first solution as $\Psi_+$ and the second one as $\Psi_-$. In order to have two linear independent solutions we also
require $A_0 B_0\neq 1$.

The procedure outlined above is adopted twice: once for $A_\infty$ and $B_\infty$ and once for $A_H$ and $B_H$.
Accordingly, we perform four numerical integrations: two from $r_\star^{(1)}$ and two from $r_\star^{(2)}$ up to
$r_\star=0$ and we obtain $(\Psi_\pm^\text{right}(r_\star),\Theta_\pm^\text{right}(r_\star))$ and
$(\Psi_\pm^\text{left}(r_\star),\Theta_\pm^\text{left}(r_\star))$ respectively. Finally we construct a linear
combination of solutions:
\begin{equation}
\left(\begin{array}{c}
  \Psi_0\\
  \Theta_0
\end{array} \right) = \left\{
\begin{array}{ll}
\displaystyle
a\,\left(\begin{array}{c}
  \Psi_+^\text{right}\\
  \Theta_+^\text{right}
\end{array} \right)+b\left(\begin{array}{c}
  \Psi_-^\text{right}\\
  \Theta_-^\text{right}
\end{array}\right) \,, & r_\star>0 \,,\\
\\
\displaystyle 
c\,\left(\begin{array}{c}
  \Psi_+^\text{left}\\
  \Theta_+^\text{left}
\end{array} \right)+d\left(\begin{array}{c}
  \Psi_-^\text{left}\\
  \Theta_-^\text{left}
\end{array}\right)
\,, & r_\star<0 \,,
\end{array}
\right.\nonumber
\end{equation}
and we choose $a,\,b,\,c,\,d$ in order to satisfy the continuity conditions, Eqs.~(\ref{eqcont}). The net result of this
procedure is a set of solutions $\{\Psi_0(r_\star),\Theta_0(r_\star)\}$ which have the correct boundary conditions and
which are continuous everywhere with $\Psi_0'$ also continuous everywhere. The discontinuity
$\left[\left[\Theta_0'\right]\right]\neq0$ is related to the choice of a trial eigenvalue $\omega_0$, which is not the
\emph{correct} eigenfrequency.

Let us now denote with $\omega_1$ the correction to the trial eigenvalue, i.e. $\omega=\omega_0+\omega_1$. If $\omega_1$
is a small correction, i.e. $\omega_1\ll\omega_0$, then \cite{Watanabe:1980}
\begin{equation}
\omega_1=\frac{\mu_0(0)\left[\left[\Theta_0'\right]\right]}{\int dr_\star\!\left[\lambda_0\!\left(\frac{\partial
        P}{\partial\omega_0}\Psi_0\!+\!\frac{\partial R}{\partial\omega_0}\Theta_0\!\right)\!+\!\mu_0\!\left(\frac{\partial
        Q}{\partial\omega_0}\Theta_0\!+\!\frac{\partial S}{\partial\omega_0}\Psi_0\!\right)\right]}\label{correction}
\end{equation}
where we have defined $P(r_\star)=-\omega^2+V_{11}$, $Q(r_\star)=-\omega^2+V_{22}$, $R(r_\star)=V_{12}$,
$S(r_\star)=V_{21}$. In our case $\partial P/\partial \omega_0=\partial Q/\partial \omega_0=-2\omega_0$ and $\partial
R/\partial \omega_0=\partial S/\partial \omega_0=0$. Moreover in the Eq.~(\ref{correction}) $\lambda_0$ and $\mu_0$ are
the solutions of the conjugate equations of Eqs.~(\ref{eqm1})-(\ref{eqm2})
\begin{eqnarray}
\lambda_0''(r_\star)+(\omega^2-V_{11})\lambda_0(r_\star)&=&V_{21}\mu_0\,,\\
\mu_0''(r_\star)+(\omega^2-V_{22})\mu_0(r_\star)&=&V_{12}\lambda_0\,.
\end{eqnarray}
The correction (\ref{correction}) has been computed in Ref. \cite{Watanabe:1980} for the case of ``bound-state like''
boundary conditions. Interestingly enough, it is also valid for the more general case of boundary conditions defined in
Eq.~(\ref{BC}). In fact it is straightforward to show that contributions to Eq.~(\ref{correction}) arising from boundary
conditions (\ref{BC}) cancel each others out, if the same boundary conditions are also imposed on $\lambda_0$ and
$\mu_0$. Therefore Eq.~(\ref{correction}) can be used in an iteration scheme until we reach the required accuracy.  We
find that convergence usually occurs, within the required precision (typically
$\left|\left[\left[\Theta_0'\right]\right]/\Theta_0'(+0)\right|<10^{-6}$), in less then 50 iterations.  However, we
cannot find the entire QNM spectrum using this method. Indeed even the single equation version of this method fails to
find first overtones of Schwarzschild BHs in General Relativity \cite{Chandrasekhar:1975zz}. This is the reason why, as
discussed in the next section, we can find QNMs with this approach only for $M^4\beta\gtrsim0.5$. For smaller values of
$\beta$ the iteration scheme ceases to converge.
\section{Numerical results}\label{results}
In this Section we present the results of our numerical integrations, performed using both the time domain approach and
the iteration scheme approach in the frequency domain. The results for time domain evolutions refer to Gaussian initial
data, with a Gaussian wave-packet characterized by $v_c = 10.0$ and $\sigma = 1.0$ in Eq.~\eqref{gauss}; the field is
extracted at $r_{\star} = 50.0M$.

\subsection{Small $M^4 \beta$ limit}\label{smallbeta}

For small values of $M^4\beta$ ($\lesssim0.5$), the perturbative dynamics is characterized by a stable
exponential mode phase. The intermediate late time evolution is dominated by
\begin{equation}
\Phi(t,r_{\rm fixed}) = e^{\omega_{\rm no} \, t}
\left(\begin{array}{c}
  a\\
  b
\end{array} \right)
\end{equation}
with $\textrm{Re}[\omega_{\rm no}] = 0$ and $\textrm{Im}[\omega_{\rm no}] < 0$.

\begin{figure}[htb]
\begin{center}
\begin{tabular}{c}
\includegraphics[scale=0.36,clip=true]{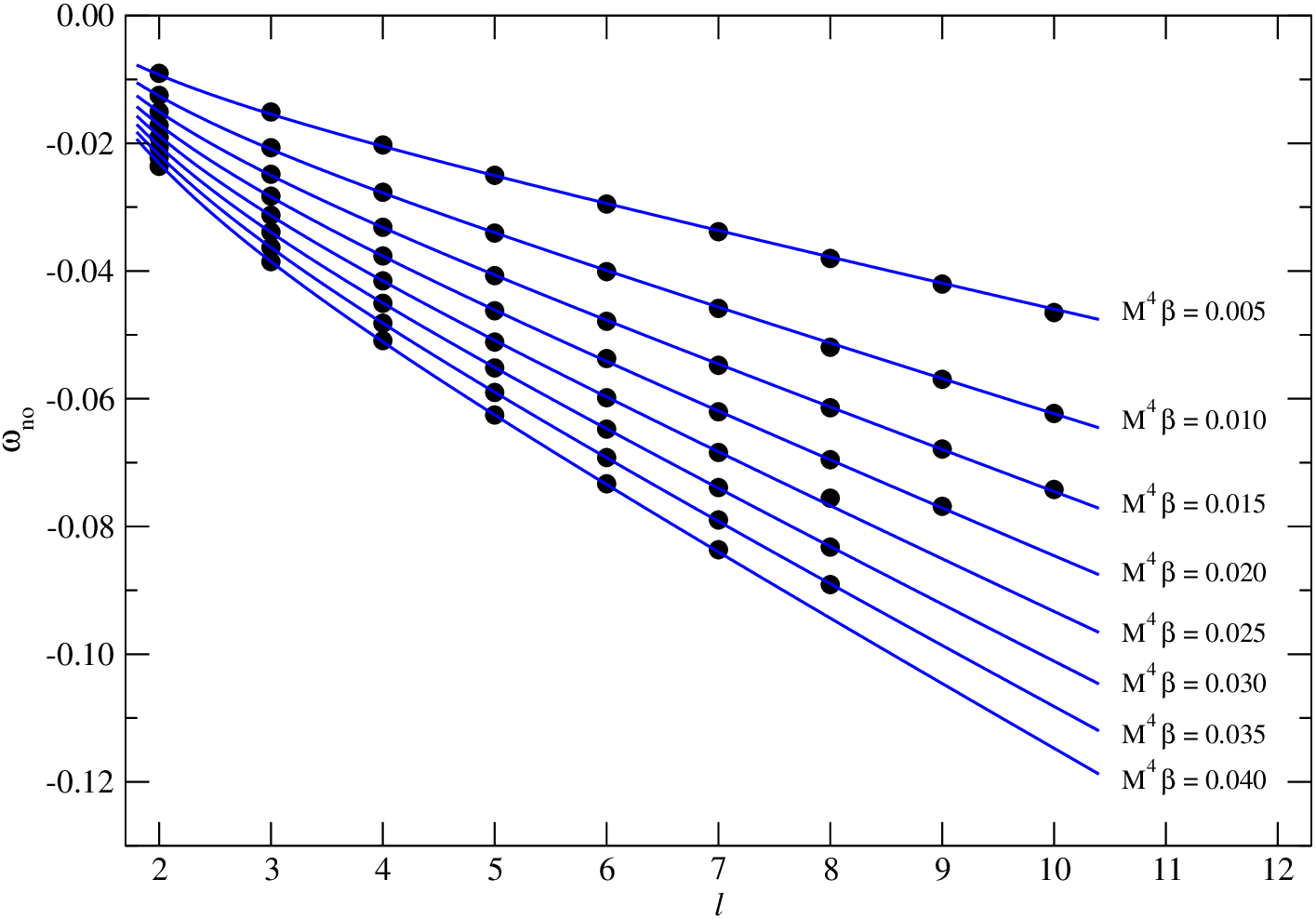}
\end{tabular}
\end{center}
\caption{\label{fig:no-mode} (Color online) $\omega_{\rm no}$ as a function of $\ell$ for different values of $M^4 \beta\le0.4$. The dots
  indicate data from our numerical methods, continuous lines indicate the fit (\ref{wno}).}
\end{figure}

After an extensive numerical exploration performed using the time domain approach, the non-oscillatory frequencies
$\omega_{\rm no}$ obtained are consistent with the expression
\begin{equation}
M\omega_{\rm no} = - 0.04024 (M^4 \beta)^{0.44} \ell 
\left(1 + \frac{2.0953}{\ell} - \frac{3.4460}{\ell^2} \right)\,,
\label{wno}
\end{equation}
which is illustrated in Fig. \ref{fig:no-mode}.

\subsection{Intermediate values of $M^4 \beta$}\label{intermbeta}
\begin{figure}[htb]
\begin{center}
\begin{tabular}{c}
\includegraphics[scale=0.36,clip=true]{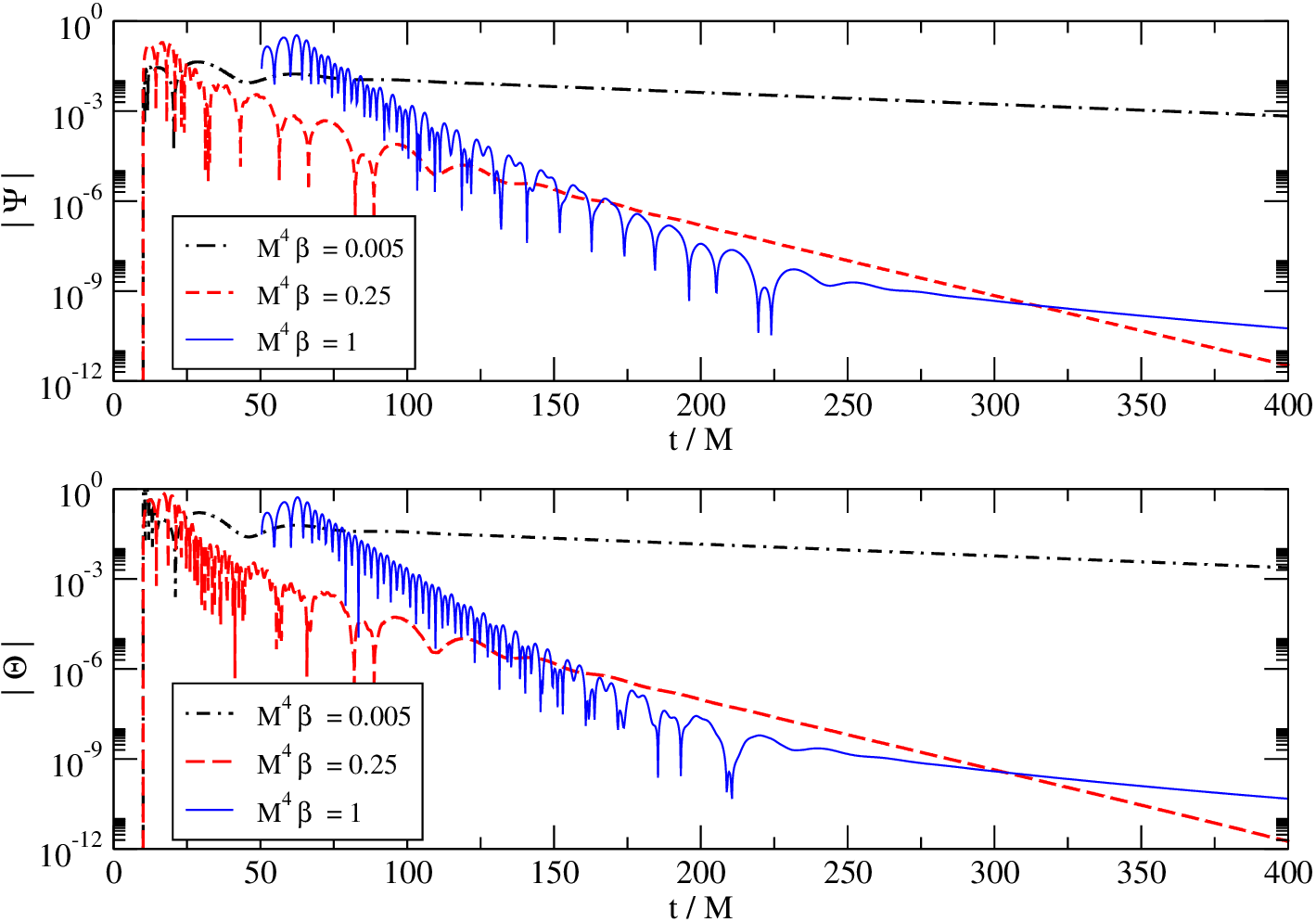}
\end{tabular}
\end{center}
\caption{\label{fig:TP1} (Color online) Time-profiles for the $|\Psi|$ (upper panel) and $|\Theta|$ (lower panel) components with
  $\ell=2$ and $M^4 \beta=5\!\cdot\!10^{-3},0.25,1$.}
\end{figure}

For $M^4\beta\gtrsim0.5$, the system evolves with damped oscillations. The transition between non-oscillating and
oscillatory mid-late time behavior can be seen in Fig. \ref{fig:TP1}, where we show the time evolution of the $\Psi$ and
$\Theta$ components with $\ell=2$ for $\beta=5\!\cdot\!10^{-3},0.25,1$ . The behavior for higher values of $\ell$ is
qualitatively similar.
\begin{table}[hbt] \centering \caption{Quasinormal frequencies for the oscillatory modes with several values of
    $M^4\beta$ and $\ell=2$. We compare the results obtained with the time domain (TD) approach with those obtained with
    the frequency domain (FD) approach.}
\vskip 5pt 
\begin{tabular}{@{}cc|c@{}}
  \hline \hline
&\multicolumn{2}{c}{$M\omega,\,\ell=2$}
\\ \hline
$M^4\beta$  &TD   &FD      \\
\hline \hline
0.50       &0.276 - 0.0967 \ii	&0.276 -0.0936 \ii   \\
           &1.98 - 0.145 \ii    &1.97 - 0.144 \ii     \\
\hline
1.00       &0.291 - 0.0970 \ii  &0.292 -0.0971 \ii   \\
           &1.43 - 0.142 \ii 	&1.43 - 0.142 \ii   \\
\hline
10.0       &0.340 - 0.0980 \ii 	&0.340 - 0.0983 \ii   \\
           &0.634 - 0.110 \ii 	&0.634 - 0.110 \ii   \\
\hline
100        &0.366 - 0.0921 \ii 	&0.367 - 0.0919 \ii   \\
           &0.501 - 0.0952 \ii 	&0.501 - 0.0954 \ii   \\
\hline
$\infty$   &0.374 - 0.0890 \ii  &0.374 - 0.0890 \ii  \\
           &0.484 - 0.0967 \ii	&0.484 - 0.0967 \ii   \\
\hline \hline
\end{tabular}
\label{tab:qnms}
\end{table}

In this oscillatory regime we have found, for each value of $M^4\beta$, two modes. In Table \ref{tab:qnms} we present
the corresponding QNM frequencies (for $\ell=2$), computed using both numerical methods described above; we find that
the agreement between the two approaches is always better then $0.4\%$. As we discuss in Section \ref{largebeta}, these
two modes belong to two different branches, which we term ``gravitational'' and ``scalar''; thus we can consider them as
the ``fundamental'' modes, i.e. the lowest lying modes of these two branches. We stress that these names refer to the
large $\beta$ limit of the modes, but both perturbations, $\Psi$ and $\Theta$, oscillate with both modes\footnote{This
  happens for $M^4\beta\lesssim 100$; for larger values
  of $M^4\beta$, the scalar perturbation $\Theta$ oscillates with one mode only, as discussed in Section
  \ref{largebeta}.}.

\begin{figure}[htb]
\begin{center}
\begin{tabular}{c}
\includegraphics[scale=0.36,clip=true]{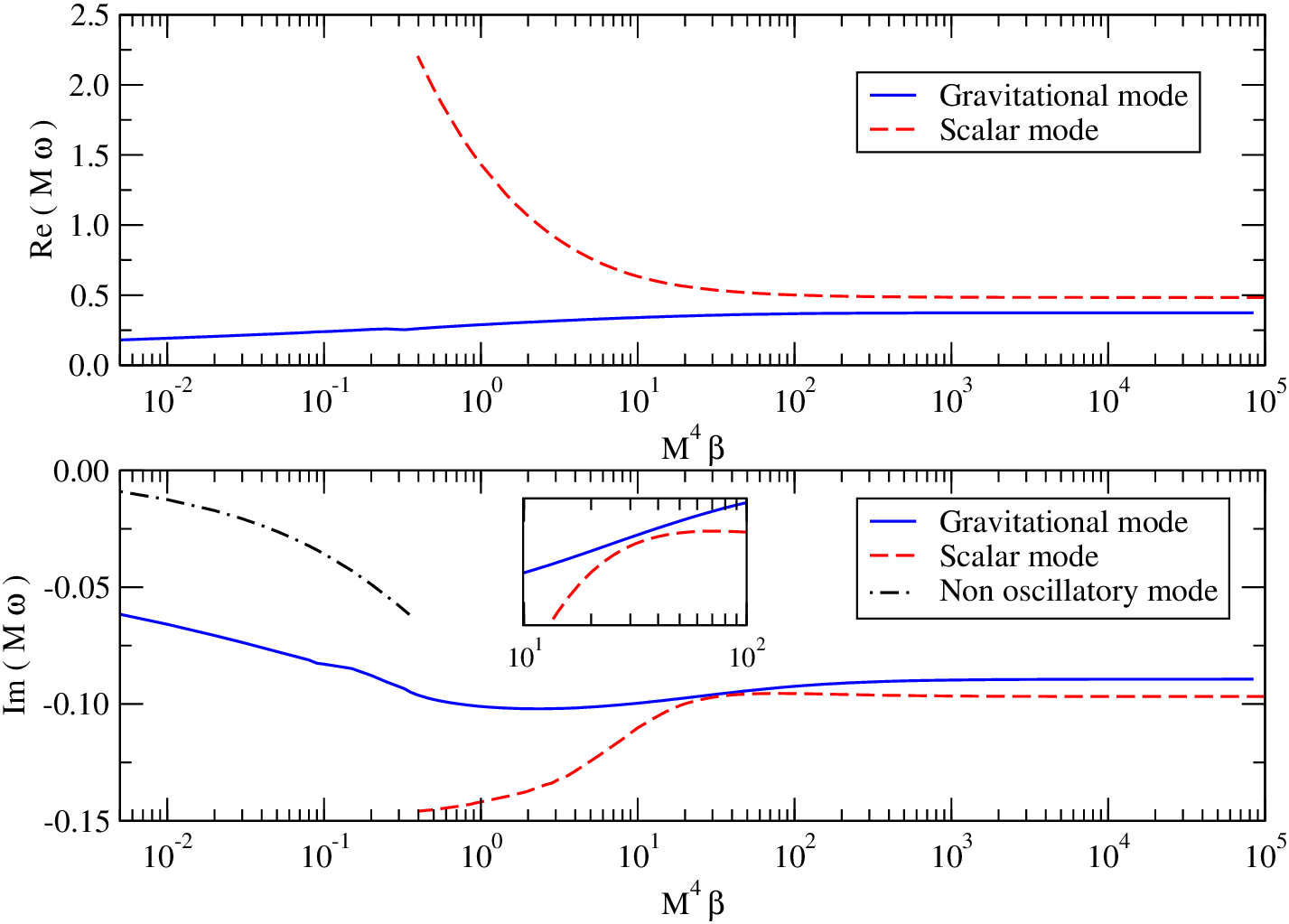}
\end{tabular}
\end{center}
\caption{\label{fig:beta_dep}  (Color online) Real (upper panel) and imaginary (lower panel) parts of the fundamental QNMs as functions
  of $\beta$ for $\ell=2$.}
\end{figure}

\begin{figure}[htb]
\begin{center}
\begin{tabular}{c}
\includegraphics[scale=0.36,clip=true]{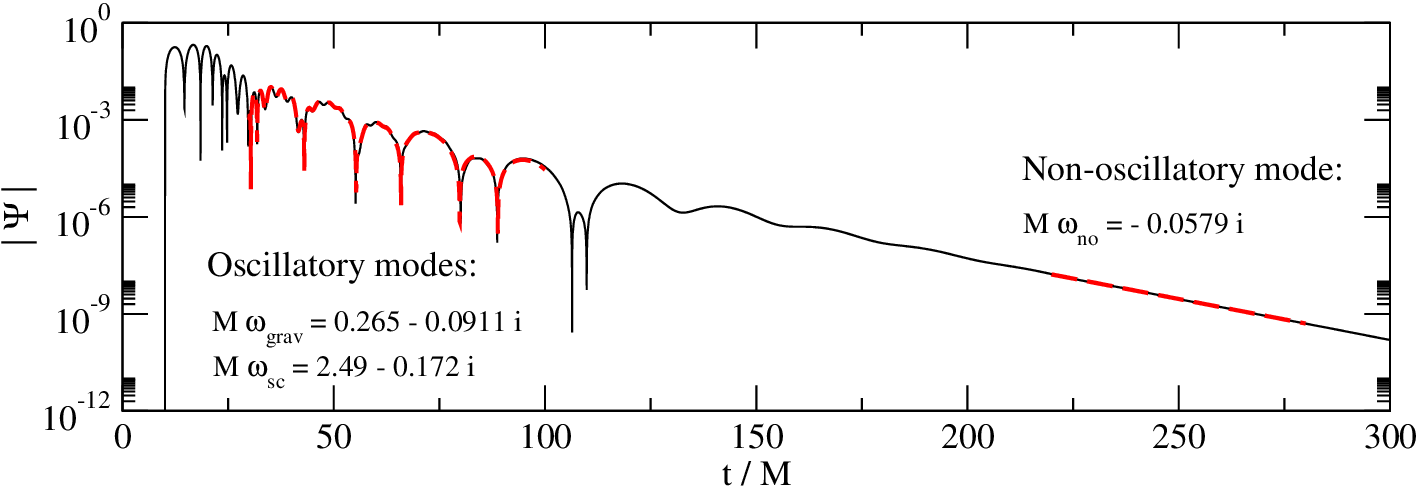}
\end{tabular}
\end{center}
\caption{\label{fig:3modes} (Color online) Time profile for the $\Psi$ component from the time-evolution approach (solid line) for
  $M^4\beta=0.3$ and $\ell=2$, compared with a combination of oscillatory and non-oscillatory modes (dashed line).}
\end{figure}

The three different $\ell=2$ modes are shown, for $10^{-2}\lesssim M^4\beta\lesssim10^5$, in Fig.~\ref{fig:beta_dep},
where the dotted-dashed line refers to the non-oscillating mode, the continuous line to the ``gravitational''
oscillating mode, and the dashed line to the ``scalar'' oscillating mode. We can see that, for small values of $\beta$,
the non-oscillating mode $\omega_{\rm no}$, which dominates the time profile, is excited together with the gravitational
oscillating mode; for $\beta=0.3$ all three modes are present, and for larger values of $\beta$ the two oscillating
modes are present. Qualitatively similar plots can be found for also for $\ell=3$ and $\ell=4$.  The time evolution of
$\Psi$ for $M^4\beta=0.3$, which is a combination of the three modes, is shown in in Fig. \ref{fig:3modes}.

It should be mentioned that the numerical determination of the different modes for the same value of $M^4 \beta$ is not
an easy task. For instance, neither of the two approaches is able to find the scalar non-oscillating mode for
$M^4\beta\sim 0.5$.  The numerical difficulties are related to the fact that the convergence of the iteration scheme in
the frequency domain approach is more difficult for small values of $\beta$. On the other hand, the time-profiles are
usually available for all the $\beta$ range considered, but the extraction of the frequencies from them is not always
possible. However, we remark that the concordance of the two methods is very good in a wide range of parameter space.

\subsection{Large $M^4 \beta$ limit\label{sec:large}}\label{largebeta}
\begin{figure}[htb]
\begin{center}
\begin{tabular}{c}
\includegraphics[scale=0.36,clip=true]{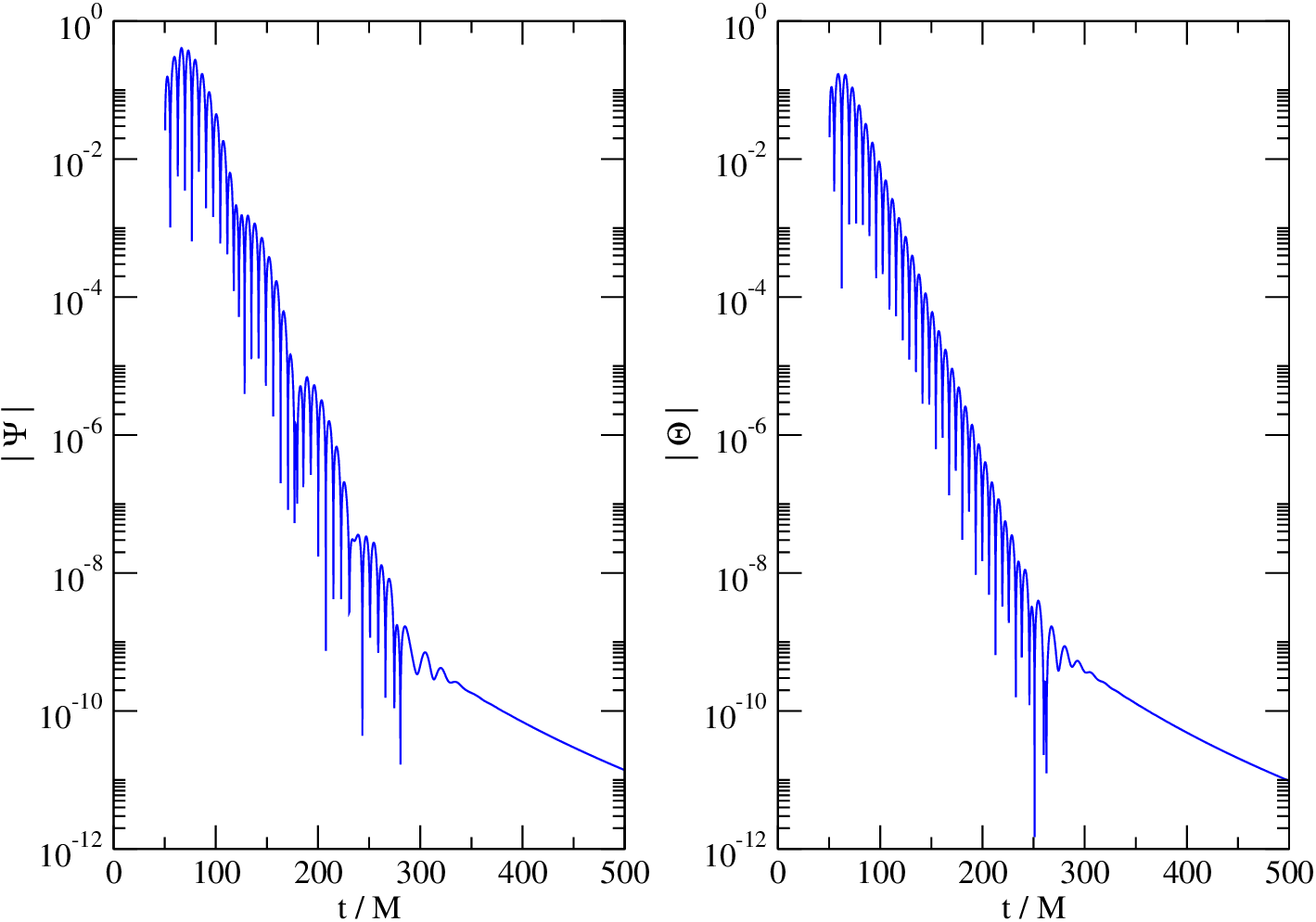} 
\end{tabular}
\end{center}
\caption{\label{fig:TP_beta100} (Color online) Time evolution of $|\Psi|$ and $|\Theta|$, for $M^4\beta=100$ and $\ell = 2$.}
\end{figure}

A time-profile for the wave function for $M^4\beta=100$ and $\ell=2$ is presented in Fig.~\ref{fig:TP_beta100}.
The data for the $\Psi$ component are consistent with a two-mode fit. The values obtained are
\begin{eqnarray}
M \omega_{\rm grav}&=&0.3736 - 0.08899\,\ii\\
M \omega_{\rm sc}&=&0.4837 - 0.09671\,\ii\,,
\end{eqnarray}
which coincide, up to numerical precision, with the complex frequencies of the (lowest lying) QNMs of Schwarzschild BHs
in Einstein's theory for gravitational ($\omega_{\rm grav}$) and scalar ($\omega_{\rm sc}$) perturbations
\cite{Berti:2009kk}. The data for the $\Theta$ component, instead, is consistent with a single mode fit, with frequency
$\omega_{\rm sc}$. The obtained frequencies fit the numerical data very accurately. We can conclude that in the
$\beta\rightarrow\infty$ limit and for low multipole numbers $\ell$, the gravitational perturbations and the scalar
field oscillate with the QNMs of Schwarzschild BHs: the former, with a combination of the scalar QNM and of the
gravitational QNM; the latter, with the scalar QNM.
This behavior can be easily understood if we consider the  $\beta \rightarrow \infty$ limit of the perturbation equations:
\begin{eqnarray}
\frac{d^2}{dr_\star^2}\Psi+\left\{ \omega^2-f\left[\frac{\ell(\ell+1)}{r^2}-\frac{6M}{r^3}\right]\right\}\Psi&=&
\frac{96\pi\,Mf}{r^5} \Theta\,,\nonumber\\
\label{eqq3}\\
\frac{d^2}{dr_\star^2}\Theta+\left\{ \omega^2-f\left[\frac{\ell(\ell+1)}{r^2}
+\frac{2M}{r^3}\right]\right\}\Theta&=&0\,.\label{eqpsi3}
\end{eqnarray}
These equations show that, as discussed in Section \ref{eqns}, the limit $\beta\to\infty$ does not correspond to the
General Relativity limit. Indeed, the gravitational field is coupled with the scalar field:
equation (\ref{eqq3}) for $\Psi$ is sourced by $\Theta$. To recover General Relativity, one should restrict to the
solutions with $\Theta\equiv0$; note that $\Theta\equiv0$ is solution of the $\beta\to\infty$ equations (\ref{eqq3}),
(\ref{eqpsi3}), not of the general equations (\ref{eqq2}), (\ref{eqpsi2}).

Eq. (\ref{eqpsi3}) coincides with the equation for scalar field perturbations of a Schwarzschild BH in General
Relativity. It does not depend on $\Psi$, and can be solved separately, yielding the well known scalar QNM frequencies
of Schwarzschild BHs \cite{Berti:2009kk}. Once Eq. (\ref{eqpsi3}) is solved, one can solve Eq. (\ref{eqq3}), treating it
like the equation of a forced oscillator, since $\Theta(r)$ can be considered as ``known''. The homogeneous equation
associated to (\ref{eqq3}) yields the gravitational QNM frequencies, like $\omega_{\rm grav}$ \cite{Berti:2009kk}, whereas
the source oscillates with frequency $\omega_{\rm sc}$. Its solution $\Psi(r)$, at very late times, oscillates with
$\omega_{\rm sc}$ only, but at earlier times it is a combination of the two frequencies, as we have found in our numerical
integrations. Furthermore, if $\Theta\equiv0$, Eq.~(\ref{eqq3}) is trivially satisfied, whereas Eq.~(\ref{eqpsi3})
simply becomes the Regge-Wheeler equation for gravitational perturbations of a Schwarzschild BH. This explains why in
the $\beta\to\infty$ limit \emph{both} scalar and gravitational QNMs are eigenfrequencies of perturbation equations.
Therefore, no matter how large the coupling constant $\beta$ is, DCS gravity leaves a peculiar signature in the
gravitational spectrum of a Schwarzschild BH. The actual detectability of this signature is discussed in Section
\ref{tests}.

\subsection{Late time power-law tails}\label{sec:tail}
\begin{figure}[htb] 
\begin{center} 
\begin{tabular}{c} 
\includegraphics[scale=0.36,clip=true]{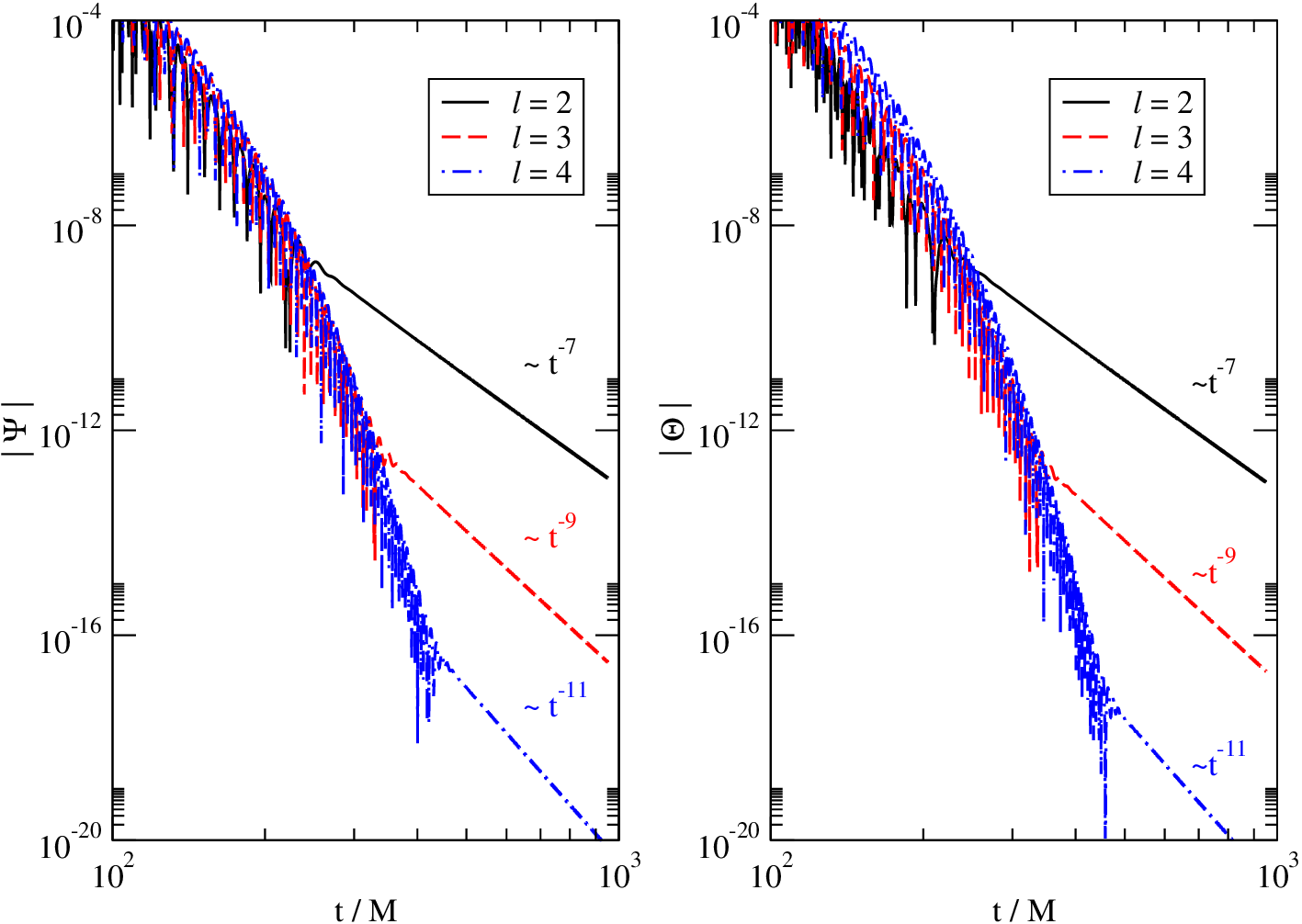}
\end{tabular} 
\end{center} 
\caption{\label{fig:latetimetails2} (Color online) Time evolution of $|\Psi|$ and $|\Theta|$ for several values of $\ell$. Straight
  lines indicate power-law decay.}
\end{figure}

Our results clearly indicate that, for large enough values of the coupling constant $\beta$, there is a power-law tail
dominating the signal of the first multipolar numbers at very late times (after the ringdown). Typical time profiles are
shown in Fig.~\ref{fig:latetimetails2}. The observed late time power-law tails are consistent with the expression
\begin{equation}
\Phi(t,r_{\rm fixed}) = t^{-(2\ell + 3)}
\left(\begin{array}{c}
  a\\
  b
\end{array} \right)\,.
\label{tail}
\end{equation}
The result (\ref{tail}) can be analytically considered in the large $r$ limit. In this limit, the equations decouple and
previous results in the literature \cite{Ching:1995tj} are applicable. The tails are universal, in sense that they show
no dependence on the parameters $M$ and $\beta$.
Note that the same behavior~(\ref{tail}) characterizes also Schwarzschild BHs \cite{CPM78}, implying that a
gravitational-wave observation of the tail would not be able to discriminate DCS gravity from General Relativity.
\subsection{Inclusion of a mass term in the Lagrangian}\label{massterm}
A relevant question is how the inclusion of a non-vanishing potential $V(\vartheta)$ in the action (\ref{action})
affects results discussed above. Here we consider the simplest potential, by including a mass term for the scalar field,
i.e.~$V(\vartheta)=m^2\vartheta^2$, with $m=G{\cal M}/(\hbar c)$ and ${\cal M}$ the physical mass of the field. We note
that, if we consider a solar mass BH, ${\cal M}=10^{-16}(mM)$ MeV; therefore, $mM=0$ for a massless field,
$10^{-13}\lesssim mM\lesssim1$ for ultra-light axions \cite{Arvanitaki:2009fg}, $mM\sim 10^{18}$ for a pion field, and
$mM\sim 10^{21}$ for a scalar field at the electroweak scale.

We note that the inclusion of a mass term destroys the shift symmetry of DCS gravity, i.e., invariance under $\vartheta
\to \vartheta + k$, with $k$ a constant. If one takes this as a fundamental symmetry, which could presumably be broken
only at the electroweak scale, it would imply that $m$ is of the electroweak size \cite{privateyunes}. Nevertheless, for
generality we do not impose any {\it a priori} constraint on the mass of the scalar field.

The mass term affects only the perturbation equation for the scalar field. In particular only $V_{22}$ in
Eq.~(\ref{V22}) is affected, and its general form for massive scalars is
\begin{equation}
V_{22}=f\left[\frac{\ell(\ell+1)}{r^2}
\left(1+\frac{576\pi M^2}{r^6\beta}\right)
+\frac{2M}{r^3}+m^2\right]\,.\label{V22mass}
\end{equation}
Note that any scalar potential $V(\vartheta)$ whose expansion for $\vartheta\ll M^2$ starts at least quadratically,
i.e. $V(\vartheta)\sim m^2\vartheta^2+{\cal O}(\vartheta^3)$, would give the same potential $V_{22}$ as
Eq.~(\ref{V22mass}). This is also the case of the periodic potential $V(\vartheta)\sim \cos{\vartheta}$ for ultra-light
axions \cite{Arvanitaki:2009fg}.

Moreover the inclusion of a mass term affects the boundary conditions (\ref{BC}) for the scalar field. In fact at infinity we have
\begin{equation}
\Theta\sim e^{i\sqrt{\omega^2-m^2} r_\star}\,.
\end{equation}
Our numerical methods are capable of computing QNMs for massive scalar perturbations whose mass is $mM\lesssim 0.2$,
which includes the case of ultra-light axions. We report that numerical results perfectly agree with our analytical
expectations in Section \ref{largebeta}. In fact, in the large $\beta$ limit, the inclusion of the potential only
affects Eq.~(\ref{eqpsi3}) and in turn it modifies only the scalar branch of modes: the QNM spectrum consists in the
usual gravitational modes plus \emph{massive} scalar modes of a Schwarzschild BH.

For smaller values of the coupling constant ($M^4\beta\lesssim 100$) the analytical limit discussed in Section
\ref{largebeta} breaks down and both gravitational and scalar modes are affected by the scalar potential. Qualitatively,
the spectrum for massive scalar perturbations is analogous to the one shown in Fig.~\ref{fig:TP1}. However, for
gravitational modes, the dependence on the scalar mass is very mild. The real part is almost insensitive to $m$ (at
least for $mM\lesssim0.2$), whereas the imaginary part changes as much as $5\%$ for $M^4\beta\sim1$ and $mM\sim0.2$.
Thus, as expected, DCS gravity leaves a signature in the QNM spectrum of a Schwarzschild BH even if a scalar potential of the form
(\ref{potential}) is included. Presumably similar results hold for larger values of $mM$ and for 
more general potentials $V(\vartheta)$.

\section{Discriminating the QNMs: no-hair tests\label{sec:nohair}} \label{tests}
Let us now consider what kind of information one can extract from gravitational-wave observations of black hole
ringdowns, i.e., from the observation of the quasinormal modes of black holes
\cite{Berti:2005ys,Berti:2007zu,Berti:2009kk}.

What we ideally would like to do is to use gravitational-wave measurements to test General Relativity and/or to rule out
alternative candidate theories. The detection of two modes in General Relativity would probably mean these modes are the
$\ell=2$ and $\ell=3$ fundamental modes, with frequencies $M\omega=0.37367-0.08896\,\ii$ and
$M\omega=0.59944-0.09270\,\ii$, respectively \cite{Berti:2009kk}. On the other hand, two-mode measurements in DCS
gravity could stand for the lowest $\ell=2$ modes, which in DCS gravity with large $M^4\beta$ are
$M\omega_{\rm grav}=0.3736 -0.08899\,\ii$ and $M\omega_{\rm sc}=0.4837 -0.09671\,\ii$. The question we now address is the
following: what minimum signal-to-noise ratio is required in order to be able to discriminate two ringdown signals, and
then to test DCS gravity? In other words, how can we tell if there really are two or more modes in the signal, and can
we resolve their parameters? If the noise is large and the amplitude of the weaker signal is very low, or the two
signals have almost identical frequencies, the two modes could be difficult to resolve. If we can resolve the two modes,
then tests of Chern-Simons predictions can be performed.

A crude lower limit on the SNR required to resolve frequencies and damping times was presented in
\cite{Berti:2005ys,Berti:2007zu,Berti:2009kk}. The analysis uses the statistical uncertainty in the determination of
each frequency and damping time, which a standard Fisher Matrix calculation estimates to be
\cite{Berti:2005ys,Berti:2007zu,Berti:2009kk},
\begin{eqnarray}
\rho \sigma_{f}&\lesssim&\frac{0.1}{M}\,,\label{sigmaffh}\\
\rho \sigma_{\tau}& \lesssim & 65M\,. \end{eqnarray}
Here, $\rho$ is the signal-to-noise ratio (SNR), $f\equiv \textrm{Re}[\omega]/2\pi$ and $\tau\equiv
1/\textrm{Im}[\omega]$ and $\sigma_k$ is the rms error for variable $k$. 
The numbers above assume white-noise for the detector, and equal amplitudes for the two modes. A
natural criterion ({\it \'a la} Rayleigh) to resolve frequencies and damping times is
\begin{equation}
\label{criterion} |f_1-f_2|>{\rm max}(\sigma_{f_1},\sigma_{f_2})\,,\qquad |\tau_1-\tau_2|>{\rm
max}(\sigma_{\tau_1},\sigma_{\tau_2})\,. 
\end{equation}
In interferometry this would mean that two objects are (barely) resolvable if ``the maximum of the diffraction pattern
of object 1 is located at the minimum of the diffraction pattern of object 2''. We can introduce two ``critical'' SNRs
required to resolve frequencies and damping times,
\begin{equation}
\rho_{\rm crit}^f =
\frac{{\rm max}(\rho \sigma_{f_1},\rho \sigma_{f_2})}{|f_1-f_2|}\,,\qquad
\rho_{\rm crit}^\tau = \frac{{\rm max}(\rho \sigma_{\tau_1},\rho
\sigma_{\tau_2})}{|\tau_1-\tau_2|}\,.
\end{equation}
We find the following estimates,
\begin{eqnarray}
\rho_{\rm crit}^f&\sim&6\,,\\
\rho_{\rm crit}^{\tau}&\sim&150\,.
\end{eqnarray}
Thus, for SNRs larger than 6, one {\it can} distinguish the two vibration frequencies in the signal, and is also able to
discriminate between the General Relativistic and the DCS prediction. For SNRs larger than 150, one can also measure and
discriminate the two different lifetimes. In other words, SNRs larger than 6 allow one to discriminate between the
$\ell=3$ ringing frequency and the ``scalar-field-type'' gravitational mode in CS gravity with large $\beta$. SNRs
larger than 150 would allow one to disentangle even the lifetime of each mode. We also note from Table \ref{tab:qnms}
and from Fig. \ref{fig:beta_dep} that for smaller values of $\beta$ the frequency of the (fundamental, ``scalar'') mode
is larger and then closer to the $\ell=3$ mode of General Relativity; to discriminate between them, a larger SNR would be
required.

The results and discussion above assume that both modes have the same amplitude. In that sense, the results above represent
a {\it lower} limit for the two modes to be discernible. In general the relative amplitude of the two modes
depends on the physical process exciting them {\it and} on the coupling parameters of the theory. 
\begin{figure}[htb] 
\begin{center} 
\begin{tabular}{c} 
\includegraphics[scale=0.36,clip=true]{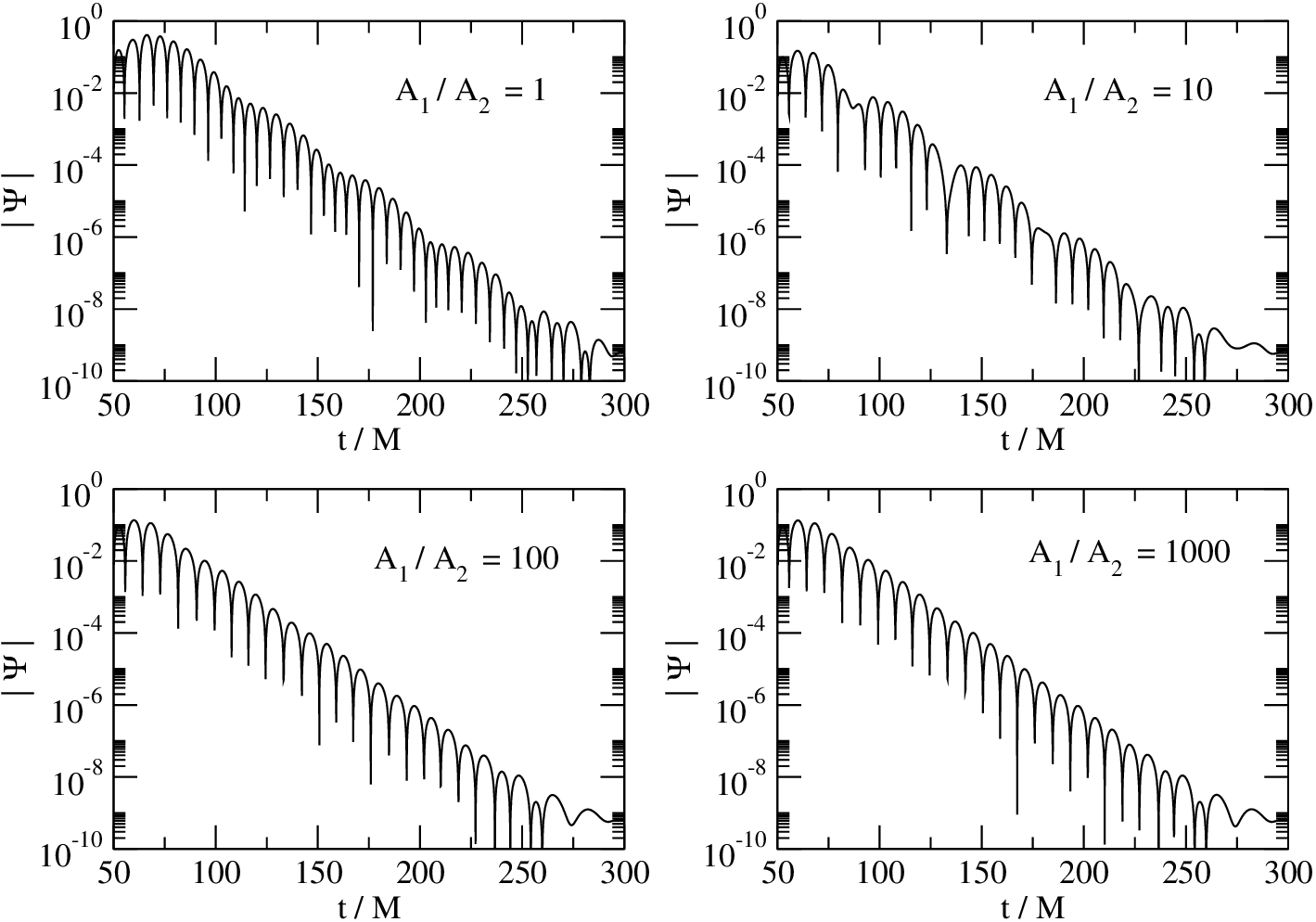}
\end{tabular} 
\end{center} 
\caption{\label{fig:dependenceinit} Dependence of the gravitational-wave signal on the relative amplitude of the initial gaussian profile,
for $\beta=100$ and $\ell=2$.}
\end{figure} 
For instance, the relative amplitude has a strong dependence on the initial amplitude of each wavepacket, as defined by equation (\ref{initdata}).
This is depicted in Fig.~\ref{fig:dependenceinit} for $\ell=2$ and $\beta=100$.
This plot shows that when $A_1/A_2=1$ the scalar and gravitational modes compete and the result is a damped beating pattern.
When $A_1 / A_2 = 1000$ the gravitational mode dominates the intermediate-time
evolution. It would be very interesting to determine the relative amplitudes of these modes for physically interesting situations, 
like extreme-mass-ratio inspirals, but this is outside the scope of the present work.

\section{Conclusions}\label{concl}
We have found that Schwarzschild BHs in DCS modified gravity are stable against axial and polar perturbations. Indeed,
an extensive investigation of BH oscillations, performed using two different numerical approaches, only yields stable
modes, either oscillating or non-oscillating.

Polar perturbations obey exactly the same master equation as in General Relativity, and therefore BHs in DCS gravity
oscillate at the same polar frequencies. Axial perturbations, instead, couple to a scalar field, enlarging the spectrum
of resonances in the gravitational sector. In particular, the ringdown of a BH in DCS gravity is a superposition of two
different QNM sectors. For large values of the constant $\beta$, which is associated to the dynamical coupling of the
scalar field, one of these sectors corresponds to the gravitational and the other sector to scalar-field QNMs of
Schwarzschild BHs in General Relativity.  Thus, a golden opportunity to test these theories is by detection of BH
ringdowns. As shown in Section \ref{sec:nohair}, a modest SNR ($\gtrsim 6$) could be sufficient to discriminate between
General Relativity and DCS modified gravity. These estimates assume very special relative amplitudes between the modes. Accurate estimates, as well as constraints on the coupling parameters, require the calculation of accurate waveforms for physically interesting processes exciting these ringdown modes.

The problem dealt with here is also interesting for a number of other reasons, in particular because we expect such kind
of problems, i.e. QNMs described by a system of coupled second order ODEs, to be a general feature of alternative and
more intricate theories; surprisingly there are very few studies of this kind of system in General Relativity.

Finally, we detail in Appendix \ref{sec:negative_beta} how ghost-instabilities develop in this theory when $\beta<0$, by
a careful analysis of the instability timescale and other features.

Generalization of our results to rotating black holes is of utmost importance, given that many astrophysical black holes are rapidly rotating.
Rotating solutions in DCS gravity are only partially understood \cite{Yunes:2009hc,Konno:2009kg}, we hope to come back to this issue in the near future.

\begin{acknowledgments} 
  We are indebted to Frans Pretorius and Nico Yunes for a careful reading of the manuscript, fruitful discussions 
  and for their many useful suggestions for improvements. We also thank Carlos Sopuerta, and 
  J.L. Costa and all the participants of the II Black Hole Workshop in Lisbon for useful comments. C.M. thanks CNPq -
  Brazil for financial support. P.P. thanks the Department of Physics, University of Rome ``Sapienza'' for the kind
  hospitality during the last stages of this work. V.C. is supported by a ``Ci\^encia 2007'' research contract and by
  Funda\c c\~ao Calouste Gulbenkian through a short-term scholarship. This work was partially supported by FCT -
  Portugal through projects PTDC/FIS/64175/2006, PTDC/FIS/098025/2008, PTDC/FIS/098032/2008, \\
  \hspace{0mm}PTDC/CTE-AST/098034/2008, CERN/FP/109290\\/2009 and by NSF grant PHY-090003 (TeraGrid). The authors thankfully acknowledge the computer
  resources, technical expertise and assistance provided by the Barcelona Supercomputing Center - Centro Nacional de
  Supercomputaci\`on.

\end{acknowledgments}

\appendix

\section{Ghost-like instabilities for $\beta<0$ \label{sec:negative_beta}}
In this Appendix we study unstable modes for the system~(\ref{eqq2}),(\ref{eqpsi2}), which arise for $\beta<0$, i.e. for
the wrong sign for the kinetic energy in the action~(\ref{action}). In particular we discuss a peculiar instability,
arising at \emph{large} multipoles $\ell$, which we believe may be seen as a general signature of ghost-like
instabilities at linear level.

For $\beta<0$, our numerical approaches both show that the amplitudes of the gravitational and scalar field grow
exponentially with time: the spacetime is unstable. The agreement between the two methods is excellent (to within the
last significant digit), thus results presented here can be reproduced by both methods.

\begin{figure}[htb]
\begin{center}
\begin{tabular}{c}
 \includegraphics[scale=0.36,clip=true]{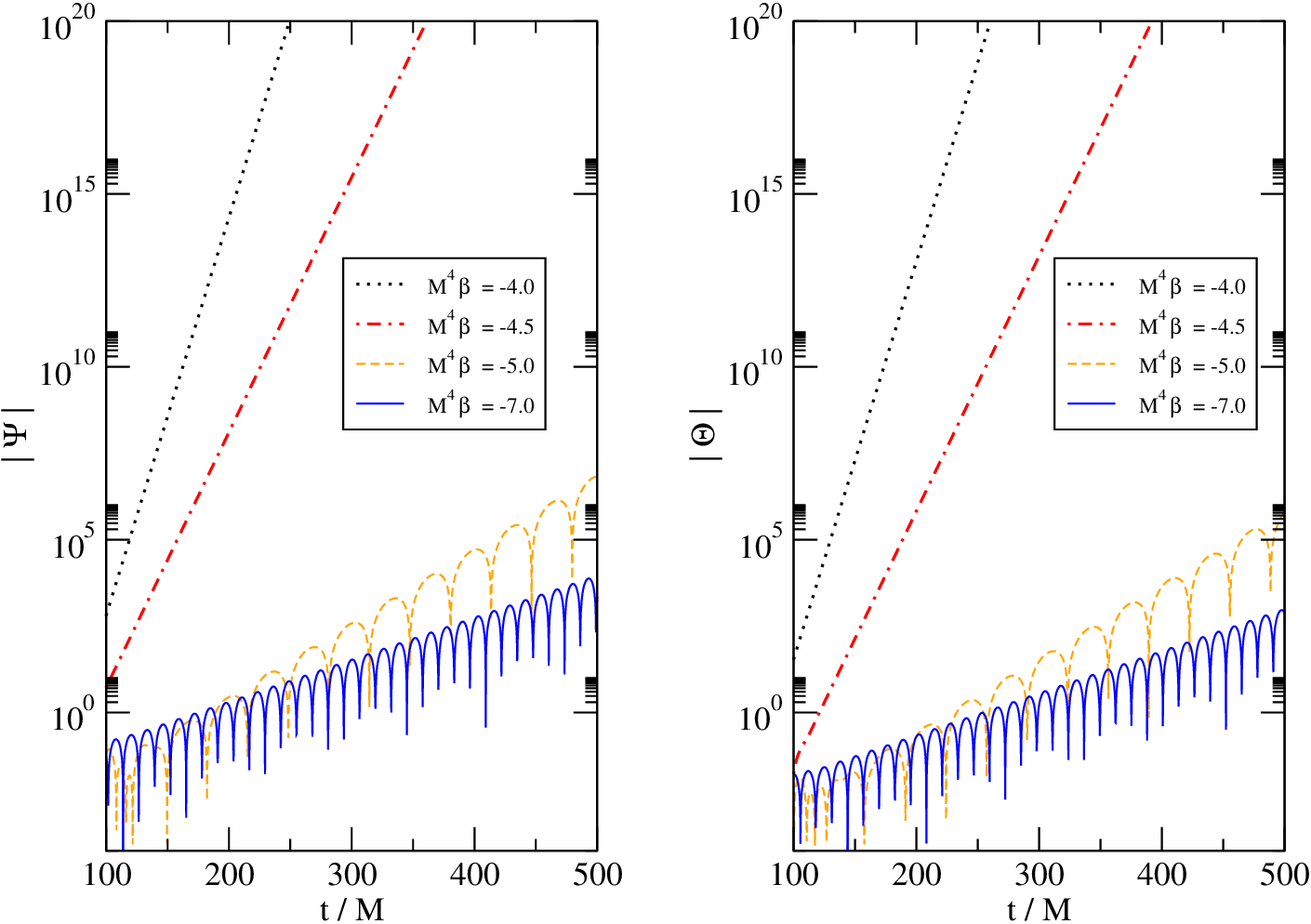}
\end{tabular}
\end{center}
\caption{\label{fig:negative-beta} (Color online) Evolution in time of $|\Psi|$ and $|\Theta|$ for 
$\ell=2$ and negative values of $M^4\beta$.}
\end{figure}

For small values of $M^4|\beta|$ the growth is purely exponential, $\sim e^{\omega_{\rm no}t}$. The non-oscillatory
exponential coefficient $\omega_{\rm no}$ depends on $\beta$ and $\ell$, as presented in Table~\ref{tab:instno}
\begin{table}[hbt] 
\centering 
\caption{Non-oscillatory exponential coefficient $M\omega_{\rm no}$ for several values of
    $M^4\beta$ and $\ell$.} \vskip 5pt \begin{tabular}{@{}cccc|cc@{}}
  \hline \hline
%
%
&\multicolumn{1}{c}{$\ell=2$}&&\multicolumn{1}{c}{$\ell=3$}&&\multicolumn{1}{c}{$M^4\beta=-1$}\\ \hline
$M^4\beta$  &$M\omega_{\rm no}$  &$M^4\beta$ &$M\omega_{\rm no}$ &$\ell$ &$M\omega_{\rm no}$\\
\hline \hline
-0.05       &5.894  &-0.05       &8.391  &2   &1.115 \\
-0.10       &4.111  &-0.10       &5.871  &3   &1.629 \\
-0.50       &1.706  &-0.50       &2.467  &4   &2.142 \\
-1.00       &1.115  &-1.00       &1.629  &5   &2.655 \\
-2.00       &0.6666 &-2.00       &0.9930 &10  &5.215 \\
-3.00       &0.4382 &-3.00       &0.6710 &20  &10.31 \\
-4.00       &0.2650 &-4.00       &0.4358 &30  &15.30 \\
-4.791      &0.0547 &-5.233      &0.0752 &50  &24.69 \\
\hline \hline
\end{tabular}
\label{tab:instno}
\end{table}
\begin{equation}
M\omega_{\rm no}\approx-0.22-0.19\ell+\frac{0.30+0.69\ell}{\left(M^4|\beta|\right)^{0.45}}\,,
\end{equation}
for any $\ell$ and small enough $M^4|\beta|$. We notice that the instability timescale $\tau=1/\textrm{Im}[\omega]$ is
shorter (the instability is stronger) for smaller $\beta$ and for larger $\ell$. From the expression above one expects
that, for large enough $|\beta|$, pure exponentially-growing modes cease to exist ($\omega_{\rm no}$ is negative for
large enough $|\beta|$). %
In fact if $\beta<-|\beta_{\rm no}|$, the late time decay is dominated by an oscillatory exponential mode. This is
depicted in Fig.~\ref{fig:negative-beta}.

Furthermore our results show that, for fixed $\ell$ and as $\beta$ further increases, there exists a critical value
$\beta_{\rm crit}$, such that
\begin{equation} 
\textrm{Im}\left[\omega(\beta_{\rm crit}) \right]=0\,,
\end{equation}
and for $\beta<-|\beta_{\rm crit}|$ the modes change from unstable to stable. This critical value depends on $\ell$ and
its dependence is very well fitted by a quadratic function
\begin{equation}
M^4\beta_{\rm crit} = -2.77 \, \ell^2 \label{fit_beta}\,.
\end{equation}
Expression (\ref{fit_beta}) implies that the complete perturbation (taking into account all multipole components) is
always unstable: for any $\beta<0$ there is always a multipole $\ell$ for which $\omega_I>0$.

In Table~\ref{tab:ldep} we present the fundamental, unstable mode, for large values of $\beta$ and different values of
$\ell$. The imaginary part of the fundamental unstable mode grows linearly with $\ell$, i.e. the instability timescale
decays linearly with $\ell$.
\begin{table}[hbt]
  \centering \caption{Fundamental unstable mode for different values of $\beta$ and taking into account the multipole
  components 
  up to $\ell=30$} \vskip 5pt
\begin{tabular}{@{}c|cc|cc|cc|cc@{}}
  \hline \hline
\multicolumn{9}{c}{Fundamental unstable mode, $\omega=\omega_R+\ii\,\omega_I$}\\
&\multicolumn{2}{c}{$M^4\beta=-10$}&\multicolumn{2}{c}{$M^4\beta=-20$}&\multicolumn{2}{c}{$M^4\beta=-30$}&
\multicolumn{2}{c}{$M^4\beta=-40$}\\ \hline
$\ell$  &$M\omega_R$   &$M\omega_I$&$M\omega_R$   &$M\omega_I$&$M\omega_R$   &$M\omega_I$&$M\omega_R$   &
$M\omega_I$\\
\hline \hline
3       &0.5387 &0.0034   	&0.5835	&0.0060 	&-	&-  		&-    	   &-\\
4       &0.7342 &0.0540   	&0.7771	&0.0371  	&0.7973 &0.0177   	&0.8074    &0.0037\\
5       &0.9154 &0.0964   	&0.9656	&0.0676  	&0.9885 &0.0443  	&1.0002    &0.0278\\
10      &1.7929 &0.2884   	&1.8862	&0.2186		&1.9251 &0.1732		&1.9456    &0.1422\\
20      &3.5266 &0.6625   	&3.7061	&0.5188		&3.7794 &0.4291		&3.8181    &0.3686\\
30      &5.2563 &1.0362   	&5.5216	&0.8186		&5.6295 &0.6848		&5.6865    &0.5947\\
\hline \hline
\end{tabular}
\label{tab:ldep}
\end{table}

Generically our results imply that for any value of $\beta<0$ there is an \emph{instantaneous} instability which
develops once all the multipolar components are taken into account. This is related to the choice of the wrong sign for
the kinetic term of the scalar field in the action, and it is the signature of ghost-like states at the linear level.


\end{document}